\documentclass[lettersize,journal]{IEEEtran}
\usepackage{amsmath,amsfonts}
\usepackage{algorithmic}
\usepackage{algorithm}
\usepackage{array}
\usepackage{booktabs}
\usepackage[caption=false,font=footnotesize,labelfont=rm,textfont=rm]{subfig}
\usepackage{textcomp}
\usepackage{stfloats}
\usepackage{url}
\usepackage{verbatim}
\usepackage{graphicx}
\usepackage{float} 
\usepackage{cite}
\usepackage{amsthm}
\usepackage{amssymb}
\usepackage{xcolor}
\usepackage{dsfont}
\usepackage{soul}
\usepackage{makecell}
\usepackage{bbding}
\newlength \figwidth
\definecolor{bittersweet}{rgb}{1.0, 0.44, 0.37}
\definecolor{glaucous}{rgb}{0.38, 0.51, 0.71}
\definecolor{gainsboro}{rgb}{0.86, 0.86, 0.86}
\definecolor{babyblueeyes}{rgb}{0.63, 0.79, 0.95}
\definecolor{silver}{rgb}{0.75, 0.75, 0.75}
\definecolor{neoncarrot}{rgb}{1.0, 0.64, 0.26}
\definecolor{Gray}{gray}{0.6}
\definecolor{LightCyan}{rgb}{0.88,1,1}
\definecolor{BackgroundLightBlue}{rgb}{0.97,0.97,1}
\definecolor{BackgroundGray}{gray}{0.98}

\newcommand{\blue}[1]{\textcolor{black}{#1}}

\def\nb0{{\mathbf{0}}}
\def\nb1{{\mathbf{1}}}

\begin{document}


\title{Constellation as a Service: Tailored Connectivity Management in Direct-Satellite-to-Device Networks}

\author{
Feng Wang,~\IEEEmembership{Member,~IEEE,}
Shengyu Zhang,~\IEEEmembership{Member,~IEEE,}
Een-Kee Hong,~\IEEEmembership{Senior Member,~IEEE,}
and Tony Q. S. Quek,~\IEEEmembership{Fellow,~IEEE
}

\thanks{Feng Wang, Shengyu Zhang, and Tony Q. S. Quek are with the Information System Technology and Design Pillar, Singapore University of Technology and Design, Singapore 487372 (e-mail: feng2\_wang@sutd.edu.sg; shengyu\_zhang@sutd.edu.sg; tonyquek@sutd.edu.sg);}
\thanks{Een-Kee Hong is with the Department of Electronic Engineering, Kyung Hee University, Yong-in 130701, Republic of Korea (email: ekhong@khu.ac.kr).}
\thanks{\emph{Corresponding author: Shengyu Zhang and Tony Q. S. Quek.}}
}



\maketitle

\begin{abstract}
Direct-satellite-to-device (DS2D) communication is emerging as a promising solution for global mobile service extension, leveraging the deployment of satellite constellations.
However, the challenge of managing DS2D connectivity for multi-constellations becomes outstanding, including high interference and frequent handovers caused by multi-coverage overlap and rapid satellite movement. 
Moreover, existing approaches primarily operate within single-constellation shell, which inherently limits the ability to exploit the vast potential of multi-constellation connectivity provision, resulting in suboptimal DS2D service performances. 
To address these challenges, this article proposes a Constellation as a Service (CaaS) framework, which treats the entire multi-constellation infrastructure as a shared resource pool and dynamically forms optimal sub-constellations (SCs) for each DS2D service region.
The formation of each SC integrates satellites from various orbits to provide tailored connectivity based on user demands, guided by two innovative strategies: predictive satellite beamforming using generative artificial intelligence (GenAI) and pre-configured handover path for efficient satellite access and mobility management. 
Simulation results demonstrate that CaaS significantly improves satellite service rates while reducing handover overhead, making it an efficient and continuable solution for managing DS2D connectivity in multi-constellation environments.

\end{abstract}


\section{Introduction}
A defining feature of 6G wireless communications is the extension of traditional terrestrial network (TN) architectures to include non-terrestrial networks (NTNs)~\cite{WigJuaSta2023}. 
This advancement aims to eliminate TN coverage gaps, ensuring ubiquitous and reliable connectivity, particularly in remote and underserved areas~\cite{GerLopBen2023}.
This transformation is enabled primarily by the large-scale deployment of cost-effective low Earth orbit (LEO) satellites. 
Tens of thousands of LEO satellites, operating at varying orbital altitudes, form multiple constellations that provide seamless and dense global coverage, supporting massive connectivity with low latency and high capacity.
This advancement also facilitates devices to  connect directly to satellites for reliable data services, representing a paradigm shift from traditional fixed satellite services to dynamic, mobile, and on-demand connectivity.

\blue{Ongoing efforts towards system-level integration of current mobile system with satellites equipped full or partial base station (BS) payloads onboard to enable direct-satellite-to-device (DS2D) cellular services within their footprints, advancing the vision of 6G ubiquitous connectivity.}
\blue{As space-terrestrial spectrum integration progresses, devices can seamlessly access satellite services without changes to hardware, firmware, or specialized applications.}
Compared to building proprietary LEO constellations, it is more practical and cost-effective for current mobile network operators (MNOs) to lease satellite resources to expand services in their underserved regions~\cite{liu2024democratizing}.
Given the increasingly crowded orbital space, this proposal is gaining widespread support by space companies and MNOs.
\blue{In this context, multiple constellations can operate concurrently~\cite{ZhouFengZhang2018}, enabling MNOs to lease specified satellites to extend coverage and address mobile service demands through robust DS2D connectivity management.}

However, efficient configuration of DS2D connectivity is challenging under multi-constellation environments, due to varying satellite coverage dynamics and DS2D traffic demands.
\blue{Under satellite multi-coverage, effective interference management is essential for reliable DS2D access, requiring more adaptive and efficient channel state information (CSI) estimation and beamforming techniques.}
Meanwhile, the coverage dynamics will cause frequent handover on DS2D devices, increasing the demand for optimized satellite mobility management strategies. 
Additionally, future mega-constellations introduce significant computational challenges, as their large scale, rapid movement, and uneven traffic distribution make real-time optimization highly complex. Managing these complexities while ensuring high-quality services necessitates innovative frameworks that can scale effectively with system demands. 
\blue{Existing studies often focus on connectivity management and resource optimization within a single satellite constellation shell~\cite{CheGuoMen2024,ZhouShengLi2023}. 
While bringing improvements, it inherently limits the vast potential of multi-constellation configurations, leading to suboptimal performance.} 
Therefore, DS2D connectivity management in multi-constellations requires more granular and strategic approaches.

To address the challenges of DS2D connectivity management in multi-constellation environments, we propose Constellation as a Service (CaaS), a novel framework that virtualizes all satellite constellations into a shared resource pool, enabling centralized control and dynamic connectivity allocation.
\blue{CaaS divides the global area into regions based on satellite traffic density, each managed by a regional MNO controller.
Leveraging predictable satellite orbits, these controllers forecast regional coverage snapshots to response to satellite movement.}
MNOs can opportunistically select the most suitable satellites from multiple constellations to form sub-constellations (SCs) for service regions. 
Within each SC, DS2D connectivity configurations are dynamically adjusted using advanced access and mobility management techniques. 
Specifically, we integrate generative artificial intelligence (GenAI) to revolutionize DS2D CSI estimation and satellite beamforming, enabling adaptive interference mitigation under multi-satellite coverage.
We also propose a pre-configured satellite handover (HO) mechanism leveraging predictable orbital trajectories to ensure a continuable quality of service (QoS) while reducing HO frequency and signaling overhead.
\blue{Guided by this integrated short- and long-term strategy, CaaS supports scalable and resilient DS2D connectivity across dynamic multi-constellation systems. 
These capabilities ultimately empower MNOs to optimize SC configurations, enabling flexible and efficient DS2D service deployment.} 
The key contributions of this article include:
\begin{itemize}
    \item We propose Constellation as a Service (CaaS), a novel satellite service provisioning framework that optimizes DS2D connectivity in multi-constellation environments.

    \item We design an advanced satellite access and mobility management strategy to enhance CaaS connectivity control, ensuring reliable and continuable DS2D services.

    \item We present CaaS operational workflow, using connectivity management insights to dynamically allocate constellation resources and  customize SC for each DS2D service region. 
\end{itemize}


\begin{figure*}[!t]
\centering
\includegraphics[width=1.4\figwidth]{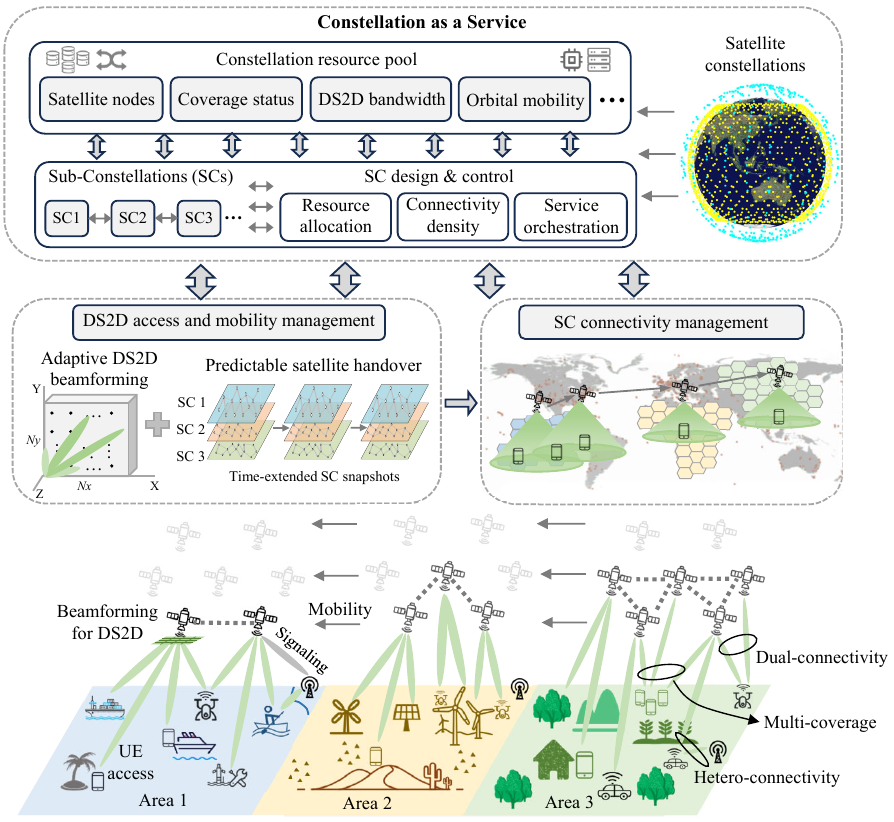}
\caption{Architecture of proposed constellation as a service (CaaS) in multi-constellation environments.}
\label{fig1}
\end{figure*}

\section{Constellation as a Service Framework}

In recent years, the industry has been actively building multiple  constellations and promoting direct satellite to device connectivity.
In parallel, the 3rd generation partnership project (3GPP) has steadily advanced DS2D standardization, evolving from Release 15 which explored satellite-based connectivity, to enabling 5G NR on satellite, and now to Release 19 which focuses on DS2D service enhancements and TN-NTN unification~\cite{SaadTarKhan2024}. 
To enable efficient DS2D connectivity management in multi-constellations, we propose the CaaS framework, as illustrated in Fig.~\ref{fig1}.
It consists of three layers: DS2D connectivity layer, onboard radio management layer, and open constellation control layer.
Each layer plays a distinct role in optimizing DS2D communication, detailed as follows.

\subsection{DS2D Connectivity Layer}

CaaS aims to provide high-speed, low-cost mobile DS2D service to users worldwide, particularly in remote and underserved areas.
In the context of 6G, DS2D connectivity is categorized into three types~\cite{TuzAguDel2023}:
\begin{itemize}
    \item \emph{Single Connectivity:} user equipment (UE) connects to a single satellite, typically in sparsely populated areas like oceanic or polar regions, or when data demand is low.
    \item \emph{Dual Connectivity:} UE under multi-coverage simultaneously connects to two satellites, either within the same or different constellations, enhancing connectivity time and throughput. 
    \item \emph{Heterogeneous Connectivity:} UE simultaneously transmits and receives data from a satellite and a ground BS for traffic load balancing, such as in docked cruise ships.
\end{itemize}
\blue{
For each UE, the connectivity type dynamically adjusts based on coverage conditions, traffic load, and user requirements.
For instance, as a UE moves from a well-served urban area to a rural region with single TN link, it may request one or more additional satellite connections when the TN link cannot meet its QoS needs.
The regional controller processes this request and assigns suitable NTN links, enabling dual connectivity to sustain service quality.
}

On the space side, satellites equipped with onboard gNB payloads and advanced processing capabilities manage UE connections within their footprint, optimizing access, mobility, and throughput~\cite{MaiCacArm2024}.
Each satellite also connects to the regional gateway and core network via either direct links or inter-satellite links (ISLs), enabling scalable and cooperative connectivity management.
Unlike terrestrial networks, DS2D connectivity management must handle large-scale concurrent connections, longer round-trip signaling, and significant Doppler shifts, increasing system complexity.
However, it benefits from the predictable satellite ephemeris information, facilitating efficient connectivity management solutions.

\subsection{Onboard Radio Management Layer}
Each satellite is equipped with extremely large antenna arrays (ELAAs), enabling high-gain beamforming to enhance DS2D connectivity.
\blue{Unlike TNs, satellite beamforming must consider rapid node movement, fluctuating channel conditions, and varying traffic demands, requiring real-time adaptability. 
AI-driven resource allocation is emerging as a key enabler for 6G NTN evolution~\cite{WuZhouLi2022}, and CaaS integrates GenAI model to optimize satellite CSI estimation and beamfroming, leveraging its powerful capability for dynamic network behavior simulation and prediction.
With multi-satellite coverage, GenAI autonomously generates optimized beamforming vectors, minimizing interference and ensuring fast and accurate onboard radio resource management.}

\blue{Additionally, traditional signal-quality-based HO mechanisms struggle in NTN due to insignificant signal strength variations across satellite cells and frequent orbital changes, leading to excessive and inefficient HOs that disrupt service continuity.}
CaaS introduces a pre-configured mobility management strategy, with a two-stage pattern:
\begin{itemize}
    \item \emph{Advanced HO Process:} For each DS2D connectivity, the HO process is divided into preparation, monitoring, and execution phases, advancing HO preparation based on predictable time of stay  and coverage changes.

    \item \emph{Pre-Configured HO Paths:} For multi-satellite coverage, UE and satellites store optimized HO sequences, allowing seamless transitions with reduced decision latency~\cite{3GPPR22210353}.  
\end{itemize}
\blue{The strategic HO path optimizes DS2D service continuity, minimizing HO failures and pin-ping HOs while lowering signaling overhead between satellites and UEs.}

\blue{While multi-satellite coverage expands access and handover options, it also amplifies interference, HO frequency, and radio access network (RAN) complexity.
CaaS addresses these by implementing coherent and adaptive satellite access and mobility management strategies, ensuring continuous high-quality DS2D connectivity in dynamic NTN environments.}

\subsection{Open Constellation Control Layer}
The open constellation control layer customizes and manages SCs for each region, comprising two key components:
\begin{itemize}
    \item \emph{Control Unit:} \blue{Consists of a central controller and regional controllers, integrated with TN control entities to capture satellite traffic and UE DS2D demands, responsible for constellation resource management and SC coordination.}

    \item \emph{SC Formation Unit:} Dynamically configures regional SCs based on the onboard radio management layer. It also interacts with the regional controller for SC updating.
\end{itemize}
All constellations are abstracted into an open resource pool, managed by the control unit to handle SC design, satellite assignment,  SC interaction, and SC lifecycle management.
Each SC is dynamically configured based on regional traffic demands:
\begin{itemize}
    \item SC size is determined by traffic density in the region.

    \item SC composition is refined based on UE requirements, selecting satellites from different constellations.
\end{itemize}
For instance, scattered UEs with low bandwidth needs are assigned to higher-altitude satellites to extend coverage duration and minimize  handovers. 
Conversely, high-throughput UE groups, such as UAV swarms, prefer very low Earth orbit (VLEO) satellites in SC to enhance latency and transmission efficiency~\cite{GioMarZor2021}.
Once configured, each SC is equipped with regional and external network interfaces to manage DS2D connectivity and HOs.

By virtualizing heterogeneous satellites into a shared resource pool, this layer enables MNOs to flexibly design SCs based on user requirements.
By fully leveraging satellites at different orbital altitudes, it ensures efficient multi-constellation resource utilization and DS2D service provision.

\section{DS2D Onboard Radio Management: Scalability and Continuity}

\begin{figure*}[!t]
\centering
\includegraphics[width=1.8\figwidth]{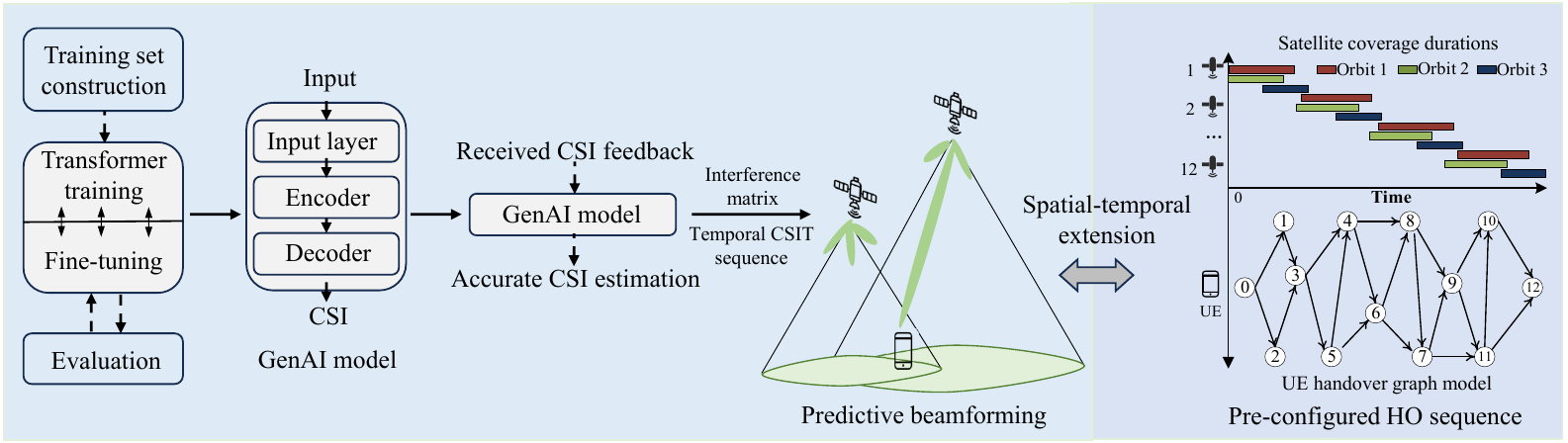}
\caption{Structure of DS2D access and mobility management: Left: GenAI-empowered satellite CSI estimation and beamforming; Right: Satellite handover graph model for pre-configured HO sequence.}
\label{fig2}
\end{figure*}

\subsection{GenAI-Empowered DS2D Access Management}

\subsubsection{Transformer-based CSI Estimation}
For DS2D connectivity in multi-constellations, heterogeneous satellites operate within the same frequency band, leading to highly dynamic CSI variations due to:
\begin{itemize}
    \setlength{\itemsep}{1pt}
    \item Satellite movement and varying atmospheric conditions.

    \item  High interference levels from multi-coverage overlapping.

    \item  Significant signal latency as long propagation distances.
\end{itemize}
To address these challenges, we employ GenAI with a transformer model for efficient and accurate CSI estimation in multi-constellation environments.
Unlike traditional deep learning models, transformer captures long-range dependencies in time-series data even in the absence of complete or up-to-date information, making it well-suited for predicting satellite CSI with time-dependent variation features.
It's self-attention mechanism dynamically assigns weights to past CSI samples, prioritizing  the most relevant historical data.
These advantages help to generate accurate CSI at the transmitter (CSIT) in dynamic NTN scenarios.
The GenAI-based CSI estimation framework, illustrated in Fig.~\ref{fig2}, consists of three key training phases:
\begin{itemize}
    \item \emph{Pre-training:} Historical CSI data sequences (channel gain, path loss, Doppler shift) are collected as the foundation. GenAI analyzes temporal patterns with dynamically weighting and simulates diverse DS2D channel conditions to enrich the training dataset.

    \item  \emph{Fine-tuning:} Model parameters are adapted to specific multi-constellation environments, improving CSI estimation accuracy.

    \item  \emph{Evaluation:} The model undergoes iterative validation to ensure robustness and generalization across varying NTN conditions.
\end{itemize}
Beyond CSI estimation, real-time interference management is achieved by leveraging predicted CSI to compute interference coefficients between transmitters and UEs, constructing an interference matrix to optimize power allocation and beamforming, minimizing inter-satellite interference.

\subsubsection{Predictive Satellite Beamforming}
A key challenge for satellite beamforming is real-time CSI processing  due to massive UE connections and inherent processing delays.
To address this, we propose a predictive beamforming strategy in satellite environments, directly generating beamforming parameters using historical CSI sequences.
Unlike traditional beamforming, which relies on CSIT feedback from the last communication round, our approach retains a fixed number of historical CSI reports and applies GenAI to predict future CSIT, generating optimized beamformers. 
In CaaS, beamforming is designed to maximize beam gain while minimizing interference, ensuring enhanced spectral efficiency.
This historical CSI estimation and beamforming approach effectively adapts to satellite channel dynamics, user distribution, and coverage variations, improving real-time satellite beamforming accuracy and DS2D rates in multi-constellation environments.


\subsection{Pre-Configured DS2D Mobility Management}

\subsubsection{Spatial Satellite Signal Distributions}
The high mobility of LEO satellites presents challenges for maintaining continuous DS2D connectivity, requiring advanced handover strategies, especially in multi-constellation environments.
Unlike terrestrial networks, DS2D signal strength experiences insignificant variations from cell center to edge due to similar propagation distances, but it varies significantly across satellites at different altitudes. 
Higher-altitude satellites (e.g. 1200\,km) provide wider coverage and longer connectivity stay time but lower signal strength, while lower-altitude satellites (e.g. 550\,km) offer stronger signals but shorter connectivity times. 
In multi-constellation scenarios, UEs may be covered by satellites from both the same and different constellations, leading to different HO options:
\begin{itemize}
    \item \emph{Same-constellation HO:} Small differences in signal strength may cause frequent and unnecessary ping-pong HOs, increasing signaling overhead.

    \item \emph{Cross-constellation HO:} The optimal target satellite is selected based on UE requirements, prioritizing service capability and stability.
\end{itemize}
In CaaS, we define two primary HO metrics: link capability and remaining time of stay.
Users can weight two metrics differently based on service demands to optimize HO benefits.
Additionally, to address varying signaling delays across different satellite altitudes, we utilize a conditional HO mechanism, which advances the HO preparation phase using predictable satellite orbital motion to process the HO request and acknowledgment, reducing signaling misalignment.

\subsubsection{Temporal Satellite Coverage Variations}
UEs will experience frequent coverage transitions in multi-constellations, leading to excessive HOs.
Additionally, HOs are not dependent due to service time connection, meaning a poor HO decision may follow a good one, creating "HO detours" that disrupt DS2D service continuity. 
To address this, we propose a pre-configured HO sequence strategy.
Specifically, we model predictable SC coverage dynamics as a handover graph model (HGM), as illustrated in Fig.~\ref{fig2} and comprised by:
\begin{itemize}
    \item \emph{Vertices:} Representing satellites in SC, ordered by coverage start time.

    \item \emph{Directed edges:} Representing feasible HOs based on coverage overlap.

    \item \emph{Weights:} Representing corresponding HO benefits based on UE-specific HO metrics.
\end{itemize}
By adding a source vertex (UE) at 0, a path from vertex 0 to the last vertex in HGM represents a possible HO sequence.
Using HGM, the source satellite computes the cumulative HO benefits for all sequences, enabling UE to pre-select an optimized HO path, 
minimizing HO detours and signaling overhead, ensuring continuous DS2D service.

\begin{figure}[!t]
\centering
\includegraphics[width=0.9\figwidth]{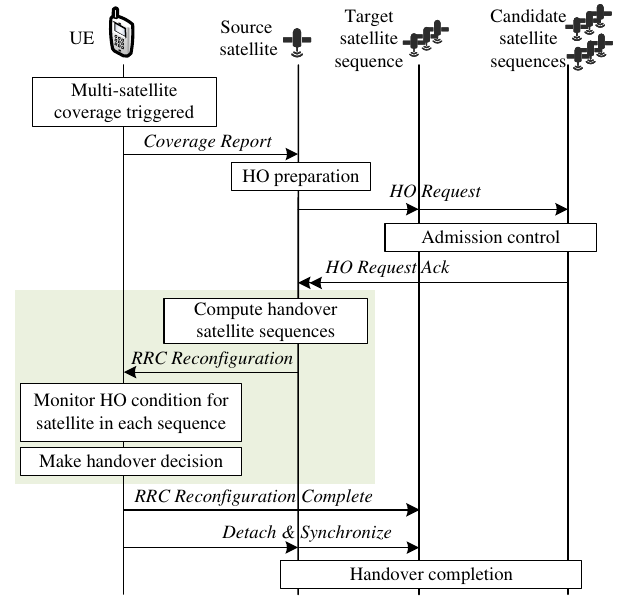}
\caption{Proposed DS2D handover procedure with pre-configured satellite service sequences.}
\label{fig3}
\end{figure}

\subsubsection{HO Procedure in Multi-constellations}
As illustrated in Fig.~\ref{fig3}, the proposed HO strategy follows a structured process.
In advanced HO preparation phase, the source satellite predicts orbital movements, sends HO requests to potential satellites, and computes optimal HO sequences.
The UE keeps monitoring HO benefits, making handover decisions accordingly.
To reduce signaling overhead, the source satellite shares the HO sequence with all involved satellites in advance, ensuring seamless transitions.
Particularly, in dual-connectivity mode, both source satellites use multipath finding algorithms in HGM to optimize HO sequences for UEs.

The presented DS2D mobility strategy minimizes HO frequency and signaling overhead. 
\blue{Moreover, as both involve onboard resource processing, pre-configured HO sequences help GenAI models collect historical CSI data efficiently.
In turn, optimized DS2D accesses refines HO sequence selection, boosting long-term satellite services.}
This integrated satellite access and mobility management approach ensures high-quality and continuous DS2D connectivity in dynamic multi-constellation environments.

\section{CaaS Operation and Performance}  

\begin{figure*}[!t]
\centering
\includegraphics[width=1.9\figwidth]{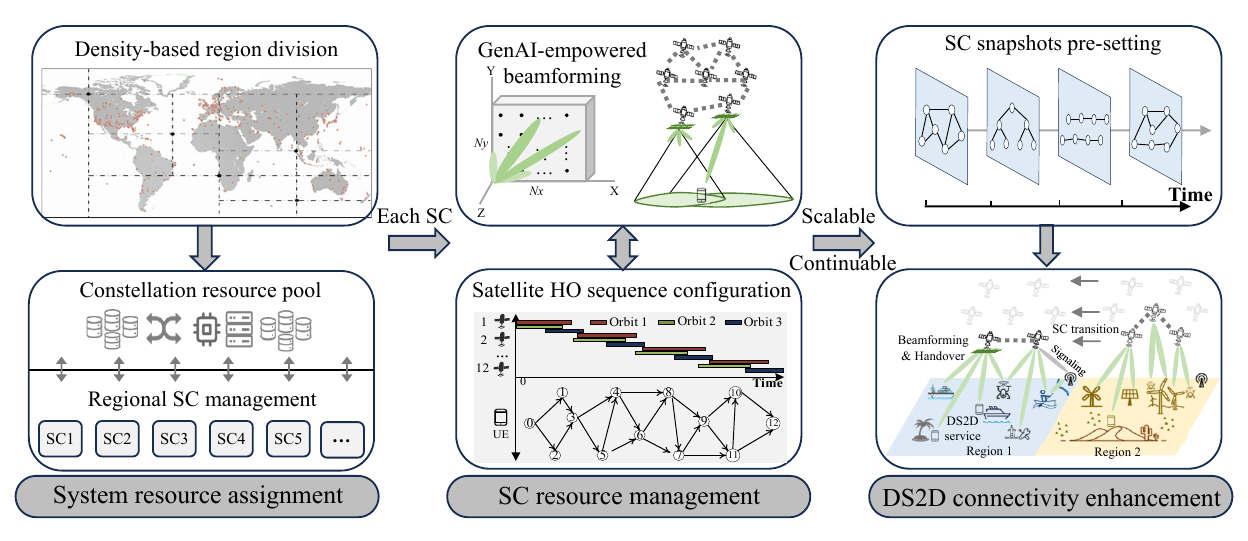}
\caption{Illustration of CaaS operation procedures with customized SC design and DS2D connectivity management for each service region.}
\label{fig4}
\end{figure*}

\subsection{CaaS Operation Procedures}
As illustrated in Fig.~\ref{fig4}, CaaS operates through a series of steps for customizing SC and DS2D connectivity management, ensuring adaptive and efficient multi-constellation resource utilization.

\subsubsection{Traffic Density Based Region Division}
To manage SCs efficiently, the global area is dynamically divided into regions based on UE traffic density. 
High traffic areas are split into smaller regions, while low traffic regions cover larger areas, ensuring that each region has a balanced connectivity management load~\cite{AlAlWang2022}.
Each region is overseen by a regional SC controller, responsible for SC management.

\subsubsection{Dynamic SC Initialization}
Each region includes satellites at different orbital altitudes, and the SC initialization for different satellite selection is based on:
\begin{itemize}
    \item UE requirements (e.g., signal strength, latency, throughput, and connectivity period).

    \item Satellite coverage and capacity (e.g., connection limits per satellite).

    \item Satellite mobility, ensuring seamless coverage for region.
\end{itemize}
This process ensures that each SC resources allocated from the constellation resource pool are sufficient for DS2D service provisioning.

\subsubsection{Optimized DS2D Connectivity}
Within each SC, advanced satellite access and mobility management strategies are applied to:
\begin{itemize}
    \item \emph{Optimize beamforming:}  Maximizing beam gain while minimizing interference, improving SC resource utilization efficiency.

    \item \emph{Enhance handover:} Pre-configuring satellite HO sequences to improve DS2D service continuity.
\end{itemize}
\blue{As UEs move across SC coverages, seamless connectivity switches occur between adjacent SCs.
By integrating these techniques, each SC enables scalable, continuable, high-performance DS2D service while minimizing signaling overhead caused by rapid satellite movement. 
Strategy results are fed back to regional controller to monitor service quality and enable timely adjustments.
To accommodate coverage changes, SCs are reconfigured over time using a series of constellation snapshots that each captures real-time coverage status and tracks its changing, ensuring seamless DS2D services. 
}

\subsection{CaaS Performance Evaluation}
In this part, we evaluate the impact of CaaS on DS2D connectivity performance, highlighting its ability to enhance multi-constellation services.
For the constellation segments, we employ the System Tool Kit (STK) simulator and consider two LEO constellations, as used by Starlink and OneWeb.
The first constellation consists of 1584 satellites distributed in 72 orbits with an inclination of 53$\mathrm{{}^\circ}$, located at altitude 550\,km.
The other constellation consists of 648 satellites distributed in 18 orbits with an inclination of 86.4$\mathrm{{}^\circ}$, located at altitude 1200\,km.
Each satellite has a 34\,dBW transmit power and 30\,dBi antenna gain, with a total 30\,MHz downlink bandwidth at 2\,GHz frequency band.
To simplify computations, we select ([0-7$\mathrm{{}^\circ}$N], [95-115$\mathrm{{}^\circ}$E]),  covering oceanic, offshore, and remote areas, as the investigated region and put up to 120 UEs in it, randomly assigning default access satellites and following a Poisson-distributed QoS demand. 
We employ a Transformer architecture for CSI prediction, with the Adam optimizer facilitating efficient parameter updates \cite{TST}.
A learning rate is initialized at 0.01 and decayed by a factor of 0.99 to ensure stable convergence and prevent overshooting the optimum. 
Moreover, each training iteration processes a batch comprising 512 sets of channel samples.
The service duration evaluated is 10\,minutes.
Under the same coverage status change, two approaches are compared:
\begin{itemize}
    \item \emph{CaaS:} Dynamically configures the optimal satellites from both constellations to form SCs and applies advanced connectivity management approaches.

    \item \emph{Standalone strategy:} Each UE remains within its default constellation, with connectivity and handover decisions based solely on signal strength.
\end{itemize}

\begin{figure}[!t]
\centering
\includegraphics[width=0.8\figwidth]{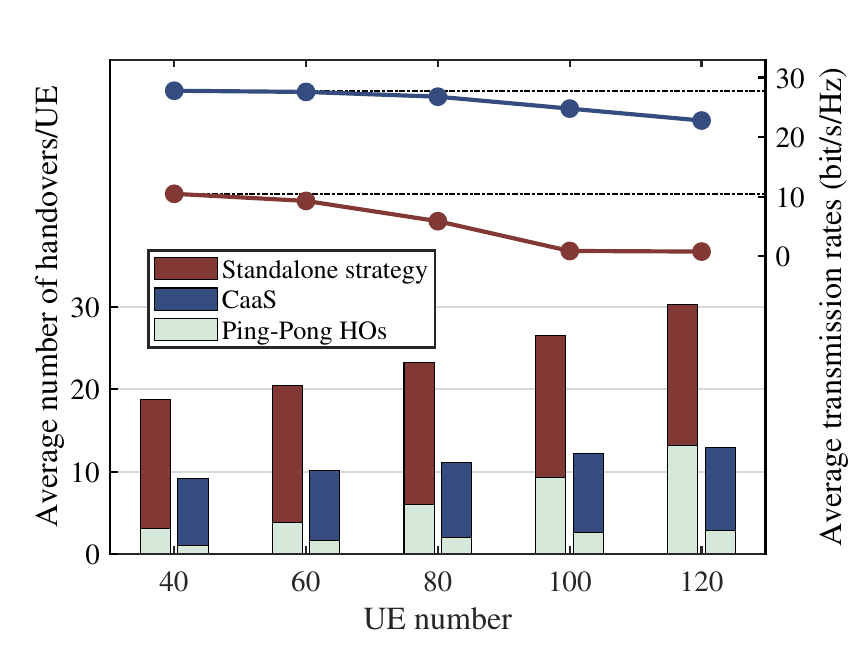}
\caption{Comparison of average transmission rates and handover frequency per DS2D connectivity between proposed CaaS and standalone strategy as the number of UEs increases.}
\label{fig5}
\end{figure}

\subsubsection*{Numerical results}
Fig.~\ref{fig5} compares the DS2D connectivity performance of two approaches under varying UE densities, evaluating average transmission rates (ATR) and HO frequency per UE to assess DS2D service capability and continuity.
Results indicate that CaaS achieves nearly three times higher ATR than the standalone strategy with 40 UEs, attributed to optimized beamforming, which dynamically assigns satellites and beams from two constellations based on signal conditions and UE demands. 
As UE density increases, CaaS effectively mitigates interference, ensuring stable ATR performance across multi-constellation coverage.
As depicted in  Fig.~\ref{fig5}, CaaS reduces HO frequency by over 50\%, optimizing handover decisions through pre-configured HO sequences. 
This significantly minimizes ping-pong HOs caused by alternating coverage between constellations, particularly in high-UE scenarios.
\blue{The results show that CaaS enhances DS2D service continuity and makes the most of satellite resources in dynamic multi-constellation environments.}

\section{Open Research Issues and Directions}
Despite the expected potential of proposed CaaS framework, there are still several challenges to be solved for DS2D services to fully unlock the benefits of satellite communications.

\textbf{TN-NTN Spectrum Sharing:}
Available spectrum is crucial for DS2D capacity and throughput.
Spectrum sharing within constellation can enhance satellite service capability by leveraging global coverage and predictable orbital movements of satellites.
\blue{Moreover, sharing spectrum between TN and NTN creates more available bandwidth and, through progressively tighter radio interface integration, enhances service capability while enabling seamless mobility for users~\cite{ChiTriKum2024}.
}

\textbf{Multi-Constellation Orchestration:}
Compared to separate single-constellation operation, a flexible DS2D connectivity architecture allows users to switch between constellations based on service needs,  requiring efficient  multi-constellation service orchestration. 
For example, delay-tolerant traffic can be assigned to MEO constellation for stability, while delay-sensitive traffic is assigned to LEO constellation for lower latency.
MEO constellation can also serve as load-balancing partners for LEO networks, optimizing satellite service distribution through task offloading.

\textbf{DS2D for Low-Altitude Economy:}
Ground BSs, typically equipped with downtitled antennas for ground users, struggle to provide reliable aerial connectivity.
As growing services for low-altitude economy, DS2D offer a promising solution for UAV connectivity, supporting critical aerial missions.
However, DS2D deployment in low-altitude environments needs to address key issues including UAV mobility support, dynamic TN-NTN segment switching, and interference mitigation in co-existed TN and NTN.

\textbf{6G Technologies for DS2D:}
\blue{6G advancements will bring novel technologies to enhance DS2D services, including  digital twin-enabled adaptive traffic scheduling for green and smart DS2D communication, reconfigurable intelligent surface for interference mitigation between TN and NTN, and onboard multicast-broadcast services for efficient group command and control.
Further research is needed to integrate and adapt these advancements into DS2D architectures, optimizing constellation operations and QoS.}

\section{Conclusions}

In this article, we explored enhancements to DS2D connectivity management in multi-constellation systems.
\blue{We begin by proposing the Constellation as a Service (CaaS) framework, which virtualizes heterogeneous constellations into a resource pool, enabling network operators to take advantage of satellites at different altitudes to customize DS2D services for different demands.}
In CaaS, the regional sub constellation (SC) configuration is driven by two key radio resource management technologies: GenAI-based satellite beamforming, which enhances interference management and spectral efficiency, and pre-configured satellite handover sequences, which optimize mobility management under multi-coverage dynamics.
These techniques allow adaptive DS2D connectivity configurations in each SC, ensuring both DS2D service capability and continuity. 
Experimental results demonstrate that CaaS can significantly increases DS2D service rates while reducing mobility overhead compared to static DS2D strategies.
Finally, we identified several aspects that need further investigation to fully unlock the potential of DS2D service in future satellite networks.


\bibliographystyle{IEEEtran}
\bibliography{main}

\section*{Biography}
\begin{IEEEbiographynophoto}{Feng Wang} received the Ph.D. degree in University of Electronic Science and Technology of China (UESTC), China, in 2022. He is now a Post-Doctoral Research Fellow at Singapore University of Technology and Design (SUTD), Singapore. His research interests include connectivity management and service orchestration for Non-Terrestrial Networks (NTN).
\end{IEEEbiographynophoto}

\begin{IEEEbiographynophoto}{Shengyu Zhang}  received the Ph.D. degree from the University of Hong Kong, Hong Kong, China, in 2023. He is currently a Post-Doctoral Research Fellow at Singapore University of Technology and Design. His research interests include networking, communication and deep learning applications.
\end{IEEEbiographynophoto}

\begin{IEEEbiographynophoto}{Een-Kee Hong} received the Ph.D. degree in electrical engineering from Yonsei University. He is currently a Professor and Dean of Research at Kyung Hee University, South Korea. He was a Senior Research Engineer with SK Telecom, and NTT DoCoMo. He served as a President of KICS. His research interests include next generation wireless communications and radio resource management.
\end{IEEEbiographynophoto}

\begin{IEEEbiographynophoto}{Tony Q. S. Quek} (S'98-M'08-SM'12-F'18) received the Ph.D.\ degree in electrical engineering and computer science from the Massachusetts Institute of Technology in 2008. Currently, he is the Associate Provost (AI \& Digital Innovation) and Cheng Tsang Man Chair Professor with Singapore University of Technology and Design (SUTD). He also serves as the Director of the Future Communications R\&D Programme, and the ST Engineering Distinguished Professor. He was honored with the 2008 Philip Yeo Prize for Outstanding Achievement in Research, the 2012 IEEE William R. Bennett Prize, the 2017 CTTC Early Achievement Award, the 2017 IEEE ComSoc AP Outstanding Paper Award, the 2020 IEEE Communications Society Young Author Best Paper Award, the 2020 IEEE Stephen O. Rice Prize, the 2022 IEEE Signal Processing Society Best Paper Award, the 2024 IIT Bombay International Award For Excellence in Research in Engineering and Technology, and the IEEE Communications Society WTC Recognition Award 2024. He is an IEEE Fellow, a WWRF Fellow, and a Fellow of the Academy of Engineering Singapore.
\end{IEEEbiographynophoto}

\end{document}